\documentclass[%
 aip,
 apl, 
 reprint,
]{revtex4-1}

\usepackage{amsmath} 
\usepackage{amssymb}

\usepackage{graphicx}
\usepackage{dcolumn}
\usepackage{bm}

\usepackage[utf8]{inputenc}
\usepackage[T1]{fontenc}
\usepackage{mathptmx}
\usepackage{etoolbox}
\usepackage{makecell}
\usepackage{physics}
\usepackage{multirow} 
\usepackage{float}
\usepackage{comment}

\makeatletter
\def\@email#1#2{%
 \endgroup
 \patchcmd{\titleblock@produce}
  {\frontmatter@RRAPformat}
  {\frontmatter@RRAPformat{\produce@RRAP{*#1\href{mailto:#2}{#2}}}\frontmatter@RRAPformat}
  {}{}
}%
\makeatother
\usepackage[colorlinks=true, linkcolor=blue, citecolor=magenta, urlcolor=magenta]{hyperref}

\begin{document}


\title{A reduced-cost two-component relativistic equation-of-motion coupled cluster method for the double electron attachment problem}
\author{Sujan Mandal}
\affiliation{Department of Chemistry, Indian Institute of Technology Bombay, Powai, Mumbai 400076, India}

\author{Tamoghna Mukhopadhyay}
\affiliation{Department of Chemistry, Indian Institute of Technology Bombay, Powai, Mumbai 400076, India}

\author{Achintya Kumar Dutta*}
\affiliation{Department of Chemistry, Indian Institute of Technology Bombay, Powai, Mumbai 400076, India}
\email{achintya@chem.iitb.ac.in}

\begin{abstract}
We present a computationally efficient relativistic formulation of the equation-of-motion coupled-cluster (EOM-CC) method for the double electron attachment (DEA) problem. In this work, the exact two-component Hamiltonian within the atomic mean-field approximation is employed, yielding results that are in close agreement with the corresponding four-component calculations. However, canonical DEA-EOM-CCSD calculations become prohibitively expensive for heavy elements and large basis sets due to the substantial memory requirements associated with the complex-valued $3p1h$ excitation manifold. To address this limitation, we introduce a new state-specific frozen natural spinor basis that significantly reduces the virtual space through two controllable truncation thresholds. Furthermore, the use of Cholesky decomposition for the two-electron integrals provides an additional reduction in memory requirements. The performance of the proposed approach is demonstrated through calculations of double ionization potentials and excitation energies for group-12 and group-14 heavy elements. Vertical excitation energies for heavy chalcogen dimers are also presented. In addition, a range of diatomic spectroscopic constants is evaluated for group-13 hydrides. Finally, the method is applied to predict the singlet-triplet gaps of dihalocarbenes, indicating that an accurate description of these systems may require excitation manifolds beyond the $3p1h$ space.
\end{abstract}

\maketitle

\section{\label{introduction}Introduction}
Since its development by \v{C}\'{\i}\v{z}ek in the 1960s, coupled-cluster (CC) theory has established itself as a preferred framework for \textit{ab initio} theoretical calculations where high accuracy and reliability are essential.\cite{cizek1966correlation,cizek1971correlation,bartlett2007coupledcluster,crawford2000introduction,shavittManyBodyMethodsChemistry2009} The method provides a hierarchical series of approximations, obtained through excitation-level truncation from a mean-field reference determinant. Modern excited-state studies in molecular systems often rely on the equation-of-motion (EOM)\cite{rowe1968equationsofmotion,geertsen1989equationofmotion,comeau1993equationofmotion,stanton1993equationb} and linear-response extensions\cite{monkhorst1977calculation, dalgaard1983aspects, mukherjee1979responsefunction, takahashi1986timedependent, koch1990coupled, koch1990excitation} of single-reference CC theory. The EOM-CC framework is appealing because it allows many electronic states to be computed from a single model Hamiltonian. Another advantage of EOM-CC approaches is that open-shell electronic states can be generated in a straightforward way by attaching or detaching electrons to a closed-shell reference state. Among the various EOM-CC techniques, the EE-EOM, IP-EOM, and EA-EOM methods are the most commonly applied for determining excitation energy (EE) spectra, ionization potentials (IP), and electron-attachment (EA) energies in molecules.\cite{rowe1968equationsofmotion, nooijen1995equation,nooijen1992coupled, hirata2000highorder,krylov2008equationofmotion} Meanwhile, the double ionization potential (DIP) and double electron attachment (DEA) variants offer an additional advantage: they provide access to both singlet and triplet manifolds, making them well-suited for studying diradicals. In the literature, there are several studies where DIPs have been computed using the non-relativistic DIP-EOM-CCSD($3h1p$) method.\cite{nooijen1997similarity,nooijen2002state,sattelmeyer2003use,demel2008application,musial2011multireference,shen2013doubly, manisha2026equationofmotion} More recently, this methodology has been extended by Piecuch and co-workers to incorporate higher-order cluster amplitudes and excitation classes.\cite{gururangan2025double} We note that multireference coupled-cluster approaches, most notably the Fock-space coupled-cluster (FSCC) method, offer an alternative route to IP, EA, EE, and DIP calculations\cite{mukherjee1989use,kaldor1991fock,lyakh2011multireferencea} and have been extended to the relativistic domain.\cite{visscherFormulationImplementationRelativistic2001,mani2011fockspace,eliav2015electronic,eliav2024relativistic,zaitsevskii2025theoretical,athanasakis-kaklamanakis2025electron} In particular, the (2,0) sector of FSCC has been shown to provide an accurate description of doubly electron-attached states.\cite{musial2012multireference,musial2013potentiala, musial2014first}

Multiply charged anions are ubiquitous in the condensed phase, where they are stabilized through solvation and other electrostatic interactions. In contrast, only a few such species are known in the gas phase, as they are typically unstable with respect to autodetachment.\cite{wang1998photodetachment,wang2000probing} Owing to this instability and the corresponding lack of reliable experimental benchmarks for DEA energies, the DEA-EOM-CCSD method has received comparatively limited attention. Nonetheless, the method is well-suited for constructing potential energy curves and for calculating excitation energies in systems where standard approaches fail.\cite{musial2011multireferenceg} In earlier non-relativistic works, DEA-EOM-CCSD has been applied to compute DIPs, EEs, and transition properties, and to construct potential energy surfaces.\cite{tomza2012optimized, shen2013doubly, shen2014doubly, ajala2017economical, gulania2021equationofmotiona, musial2014equationofmotiona, zhao2026analytic,shen2021double} Guo et al.\cite{guo2020equation} further demonstrated its applicability to heavy elements by calculating EEs with spin-orbit coupling (SOC) treated at the post-Hartree-Fock stage.

In this contribution, we present a relativistic extension of the DEA-EOM-CCSD($3p1h$) method. Relativistic effects are especially important for systems containing heavy elements, and the most rigorous way to incorporate them is through the four-component Dirac–Coulomb Hamiltonian. However, the explicit treatment of the small component substantially increases the computational cost of four-component calculations, limiting their routine use for medium- and large-sized molecules. Within the family of two-component relativistic methods,\cite{hess1986relativistic,lenthe1993relativistic,dyall1997interfacing,nakajima1999new,barysz2001twocomponent,liu2009exact,saue2011relativistic} exact two-component (X2C) Hamiltonians\cite{dyall1997interfacing,liu2009exact,kutzelnigg2005quasirelativistic,ilias2007infiniteorder,dyall2007introduction} offer an efficient low-cost alternative by retaining only the positive-energy states while preserving the dominant relativistic contributions. Among the numerous X2C variants, the X2CAMF formulation,\cite{liu2018atomic,zhang2022atomic,knecht2022exact} which incorporates atomic mean-field (AMF) spin–orbit integrals, is particularly noteworthy for its accuracy and efficiency. It delivers a balanced treatment of both scalar and spin–orbit relativistic effects at a significantly reduced computational expense. A further challenge arises from the storage of four-index two-electron integrals (ERIs), which represent a major memory bottleneck. Several approaches have been proposed in the literature to compress these ERIs into low-rank tensors, with the most widely used being density fitting (DF), resolution of identity (RI), and Cholesky decomposition (CD).\cite{beebe1977simplifications, feyereisen1993use, vahtras1993integral, werner2003fast, koch2003reduced} Unlike DF/RI approaches, which employ pre-optimized auxiliary basis sets, CD constructs the auxiliary basis functions dynamically during the decomposition. In the Davidson iterations, the $3p1h$ block of the DEA-EOM-CCSD equations imposes a substantial burden on the storage requirement. The corresponding Fock-space coupled-cluster method bypasses the problem by folding it into an effective/intermediate Hamiltonian in the $2p$ space.\cite{musial2012multireference,musial2013potentiala, musial2014first} However, an equation-of-motion treatment of the double electron attachment problem remains desirable because of its black-box nature and theoretical simplicity. A potential solution for the high scaling and large storage requirement of DEA-EOM-CCSD($3p1h$) is the reduction of the virtual space through frozen natural spinors (FNS),\cite{chamoli2022reduced} which serve as the relativistic analogue of frozen natural orbitals.\cite{lowdin1955quantum} In this work, we show that employing a state-specific variant, state-specific FNS (SS-FNS),\cite{mester2017reducedcosta, folkestad2021multilevel, manna2025reduced,mukhopadhyay2025reducedcost} results in noticeably faster convergence toward the canonical DEA-EOM-CCSD results. The SS-FNS basis is constructed using lower-level methods such as DEA-CIS(D) or DEA-ADC(2),\cite{thielen2023development} and discarding spinors with very small occupation numbers yields a significant reduction in the cost of the subsequent high-level calculations. The remainder of the paper is organized as follows. Section \ref{sec:theory} details the theoretical foundations of the relativistic DEA-EOM-CCSD($3p1h$) method and includes concise summaries of the X2CAMF Hamiltonian, CD technique, and the SS-FNS framework. Section \ref{sec:compu_details} discusses the computational setup and implementation strategies. Section \ref{sec:results} benchmarks the performance of the relativistic DEA-EOM-CCSD($3p1h$) approach by comparing it with available experimental and theoretical results for atomic and molecular systems. Concluding remarks and future directions are presented in Section \ref{sec:conclusion}.

\section{\label{sec:theory}Theory}
\subsection{The X2CAMF approximation}
The relativistic four-component Dirac-Coulomb Hamiltonian\cite{dyall2007introduction} can be expressed as
\begin{equation}
\label{eq:H_4c}
    \hat{H}^{\text{4c}} = \sum_{pq}{h^{\text{4c}}_{pq}\hat{a}_p^{\dagger}\hat{a}_q} 
    + \frac{1}{4}\sum_{pqrs}{g^{\text{4c}}_{pqrs}\hat{a}_p^{\dagger}\hat{a}_q^{\dagger}\hat{a}_s\hat{a}_r}\,.
\end{equation}
The no-pair approximation~\cite{sucher1980foundations} allows one to exclude the negative energy spectrum. When using a kinetically balanced basis,\cite{stanton1984kinetic} the two-electron part of the Dirac-Coulomb Hamiltonian in Eq.~\eqref{eq:H_4c} can be separated into spin-free (SF) and spin-dependent (SD) parts under the spin separation scheme,\cite{dyall1994exact}
\begin{equation}
\label{eq:separation}
    g^{\text{4c}}_{pqrs}
    = g^{\text{4c,SF}}_{pqrs} + g^{\text{4c,SD}}_{pqrs}\,.
\end{equation}
Owing to the localized nature of spin–orbit interactions, the spin-dependent part can be effectively described using an atomic mean-field (AMF) approximation,\cite{hess1996meanfield}
\begin{equation}
\label{eq:amf}
    \frac{1}{4}\sum_{pqrs}{g^{\text{4c,SD}}_{pqrs}\hat{a}_p^{\dagger}\hat{a}_q^{\dagger}\hat{a}_s\hat{a}_r}
    \approx \sum_{pq}{g^{\text{4c,AMF}}_{pq}\hat{a}_p^{\dagger}\hat{a}_q}\,,
\end{equation}
with the AMF integrals defined by
\begin{equation}
\label{eq:amf2}
g^{\text{4c,AMF}}_{pq}=\sum_{A}\sum_{i_A}n_{i_A}\,g^{\text{4c,SD}}_{p i_A,\,q i_A}\,,
\end{equation}
where $A$ labels the distinct atoms in the molecule, $i_A$ runs over the occupied spinors of atom $A$, and $n_{i_A}$ denotes the corresponding occupation numbers. Hence, the four-component Dirac-Coulomb Hamiltonian ($\hat{H}^{\text{4c}}$) under the AMF approximation can be written as\cite{liu2018atomic}
\begin{equation}
\label{eq:4camf}
    \hat{H}^{\text{4c}} = \sum_{pq}{h^{\text{4c}}_{pq}\hat{a}_p^{\dagger}\hat{a}_q}
    + \frac{1}{4}\sum_{pqrs}{g^{\text{4c,SF}}_{pqrs}\hat{a}_p^{\dagger}\hat{a}_q^{\dagger}\hat{a}_s\hat{a}_r}
    + \sum_{pq}{g^{\text{4c,AMF}}_{pq}\hat{a}_p^{\dagger}\hat{a}_q} \,.
\end{equation}
The four-component Hamiltonian can be mapped onto a two-component framework. The one-electron part of the four-component Hamiltonian, $h^{\text{4c}}$, can be expressed in matrix form (in atomic units) as\cite{dyall2007introduction}
\begin{equation}
\label{eq:h4cmatrix}
    h^{\text{4c}} =
    \begin{pmatrix}
    V & T \\
    T & \frac{1}{4c^2}W-T
    \end{pmatrix}\,.
\end{equation}
Here, $V$, $T$, and $W$ correspond to the nuclear potential, kinetic energy, and small-component nuclear-attraction matrices, respectively, which are given by
\begin{align}
    V_{pq}&=\left\langle p|\hat{V}|q \right\rangle \nonumber \\
    T_{pq}&=\left\langle p|\frac{\vec{p}^{\,2}}{2}|q \right\rangle \nonumber \\
    W_{pq}&=\left\langle p|(\vec{\sigma}\cdot\vec{p})\hat{V}(\vec{\sigma}\cdot\vec{p})|q \right\rangle \,.
\end{align}
An exact block-diagonalization of the matrix $h^{\text{4c}}$ in Eq.~\eqref{eq:h4cmatrix} yields the effective two-component one-electron Hamiltonian, referred to as $h^{\text{X2C-1e}}$.\cite{dyall1997interfacing,liu2009exact}
\begin{equation}
    h^{\text{X2C-1e}} = R^\dagger \left(V+X^\dagger T + TX + X^\dagger \left(\frac{1}{4c^2}W-T\right) X\right) R\,.
\end{equation}
The matrix $X$ establishes the relationship between the large- and small-component coefficients, whereas the matrix $R$ connects the large-component coefficients to the corresponding two-component wavefunction,
\begin{equation}
\label{eq:X}
    C^S=XC^L\,,
\end{equation}
\begin{equation}
\label{eq:R}
    C^L=RC^{\text{2c}}\,.
\end{equation}
When two-electron picture-change (2e-pc) contributions are neglected, the spin-free part of the Coulomb interaction, $g^{\text{4c,SF}}$, reduces to the non-relativistic two-electron integrals $g^{\text{NR}}$,
\begin{equation}
    g^{\text{4c,SF}} \approx g^{\text{NR}}\,.
\end{equation}
For each atom, the 4c-AMF integrals are transformed into the two-component form ($g^{\text{2c,AMF}}$) using the $X$ and $R$ matrices. Hence, after transformation, the overall Hamiltonian reduces to the two-component X2CAMF Hamiltonian\cite{liu2018atomic,zhang2022atomic,knecht2022exact}
\begin{equation}
\label{eqn39}
    \hat{H}^{\text{X2CAMF}} = \sum_{pq}{h^{\text{X2CAMF}}_{pq}\hat{a}_p^{\dagger}\hat{a}_q}
    + \frac{1}{4}\sum_{pqrs}{g^{\text{NR}}_{pqrs}\hat{a}_p^{\dagger}\hat{a}_q^{\dagger}\hat{a}_s\hat{a}_r}\,,
\end{equation}
where $h^{\text{X2CAMF}} = h^{\text{X2C-1e}}+g^{\text{2c,AMF}}$ is the effective one-electron operator. The primary advantage of the above Hamiltonian is that it entirely circumvents the construction of relativistic two-electron integrals, which substantially reduces the computational cost of the integral transformation.

\subsection{EOM-CC theory for double electron attachment}
In CC theory,~\cite{cizek1971correlation,bartlett2007coupledcluster,crawford2000introduction, paldusCorrelationProblemsAtomic1972, lindrothNumericalSolutionRelativistic1988, visscherFormulationImplementationRelativistic1996, visscherFormulationImplementationRelativistic2001} the ground-state wavefunction is built upon an exponential ansatz,
\begin{equation}
    \ket{\psi_{\text{CC}}} = e^{\hat{T}} \ket{\phi_0}\,,
\end{equation}
where $\ket{\phi_0}$ denotes the N-electron reference determinant, which corresponds to the X2CAMF-SCF state in the present work, and $\hat{T}$ is the cluster operator, defined as
\begin{equation}
    \hat{T} = \sum_{n=1}^{M_T} \hat{T}_n \,,
\end{equation}
with $M_T\leq N$. The operator $\hat{T}_n$ can be expressed in terms of the second-quantized creation and annihilation operators
\begin{equation}
\label{eq:fullT}
    \hat{T}_n = \sum_{\substack{i<j<\dots \\ a<b<\dots}} t_{ij\dots}^{ab\dots} \left\{ \hat{a}_a^\dagger \hat{a}_b^\dagger \dots \hat{a}_j \hat{a}_i \dots \right\} \, ,
\end{equation}
where $t_{ij...}^{ab...}$ are the cluster amplitudes, obtained by solving the projected equations
\begin{equation}
\label{eq:ampccsd}
    \left\langle  \phi _{ij...}^{ab...} |\bar{H}| {{\phi }_{0}} \right \rangle =0\,.
\end{equation}
Here, the subscripts $(i, j, ...)$ and $(a, b, ...)$ denote occupied and virtual spinors, respectively, and $\left| \phi_{ij\ldots}^{ab\ldots} \right\rangle$ are the excited Slater determinants generated from the reference. The X2CAMF ground-state CC energy is obtained as
\begin{equation}
\label{eq:eccsd}
    \bra{\phi_0} \bar{H} \ket{\phi_0} = E_{\text{CC}} \,.
\end{equation}
In Eqs.~\eqref{eq:ampccsd} and~\eqref{eq:eccsd}, $\bar{H}$ is the similarity-transformed Hamiltonian, defined as
\begin{equation}
\label{eq:Hbar}
    \bar{H}=e^{-\hat{T}}\hat{H}^{\text{X2CAMF}}e^{\hat{T}} \,.
\end{equation}
Within the EOM-CC framework,\cite{geertsen1989equationofmotion,comeau1993equationofmotion,stanton1993equationb,rowe1968equationsofmotion, nooijen1995equation,nooijen1992coupled, hirata2000highorder,krylov2008equationofmotion} the $\mu^{\text{th}}$ doubly electron-attached state\cite{gulania2021equationofmotiona,shen2013doubly} is obtained from the CC ground state $| {{\psi }_{\text{CC}}} \rangle$ as given below,
\begin{equation}
    \left| {{\psi }^{\mu}} \right\rangle = \hat{R}^\mu \left| {{\psi }_{\text{CC}}} \right\rangle \,,
\end{equation}
where $\hat{R}^\mu$ is a linear double-electron-attachment operator defined as
\begin{equation}
\label{eq:Rmu}
    \hat{R}^{\mu} = \sum\limits_{m=0}^{M_R} \hat{R}^\mu_{(m+2)p\,mh} \,,
\end{equation}
with
\begin{equation}
\begin{aligned}
\label{eq:Rmuparts}
    \hat{R}^\mu_{(m+2)p\,mh} = &\sum_{\substack{i_1<\ldots<i_m\\a<b<c_{1}<\ldots<c_{m}}}
    r^{abc_1\cdots c_m}_{i_{1}\cdots i_{m}}(\mu) \\ 
    &\times \{\hat{a}^\dagger_{a} \hat{a}^\dagger_{b} \hat{a}^\dagger_{c_{1}} \ldots \hat{a}^\dagger_{c_{m}}\;\hat{a}_{i_{m}} \ldots \hat{a}_{i_1}\}\,.
\end{aligned}
\end{equation}
The energy change associated with the double-electron-attachment process, $\omega^\mu = E_\mu-E_{\text{CC}}$, is obtained by solving the DEA-EOM-CC eigenvalue equation,
\begin{equation}
    [\bar{H}, \hat{R}^\mu ]| {{\phi }_{0}} \rangle = \omega^\mu \hat{R}^\mu | {{\phi }_{0}} \rangle\,.
\end{equation}

In this work, the cluster operator is truncated at the singles and doubles level ($M_T=2$), and the EOM-CC excitation space is restricted to the $3p1h$ subspace ($M_R=1$). The $3p1h$ block carries a substantial storage overhead in the relativistic case, where all quantities are complex-valued, making canonical calculations with large basis sets impractical for heavy elements. To circumvent this bottleneck, we employ an FNS-based truncation of the virtual space, which significantly reduces the computational cost while preserving accuracy close to the canonical limit.

\subsection{State-specific frozen natural spinors (SS-FNS)}
\label{sec:theory_ssfns}
Natural spinors\cite{chamoli2022reduced} are the relativistic counterparts of natural orbitals\cite{lowdin1955quantum} and are obtained by diagonalizing the correlated relativistic one-particle reduced density matrix (1-RDM). In the FNS scheme, the 1-RDM is constructed only for the virtual–virtual block, while the occupied spinors are retained at their SCF level. Among the various available choices,\cite{chamoli2022reduced,chakraborty2025lowcost,mukhopadhyay2025reducedcost} we employ an MP2-based FNS approach\cite{chamoli2022reduced,surjuse2022lowcost,chamoli2024relativistic,yuan2022assessing,mukhopadhyay2026reducedcost,mandal2026thirdorder,chamoli2025reduced,majee2024reduced} for the ground state. The virtual–virtual block of the 1-RDM is given by
\begin{equation}
    D_{ab}^{\text{MP2}}=\frac{1}{2}\sum_{cij} {\frac{\langle ac||ij \rangle \, \langle ij||bc \rangle}{\varepsilon_{ij}^{ac} \, \varepsilon_{ij}^{bc}}}\,,
\end{equation}
where $\varepsilon_{ij}^{ac}=\varepsilon_{i}+\varepsilon_{j}-\varepsilon_{a}-\varepsilon_{c}$ (and analogously for $\varepsilon_{ij}^{bc}$), with $\varepsilon_p$ the spinor energies.
Diagonalization of $D^{\text{MP2}}$ yields the virtual natural spinors (VNS) as the eigenvectors $V$, while the associated eigenvalues $\eta$ are their occupation numbers,
\begin{equation}
    D^{\text{MP2}}V=V\eta\,.
\end{equation}
VNS with small occupation numbers are discarded using a cutoff $\eta_{\text{crit}}$: a rectangular selection matrix $\mathcal{P}$ is defined by $\mathcal{P}_{ij}=\delta_{ij}$ if $\eta_i \geq \eta_{\text{crit}}$ and $\mathcal{P}_{ij}=0$ otherwise, and the retained VNS are collected as
\begin{equation}
\label{eq:tilde}
    \tilde{V}=V\mathcal{P}\,.
\end{equation}
The virtual–virtual block of the Fock matrix is subsequently projected onto the truncated natural-spinor basis,
\begin{equation}
    \tilde{F}=\tilde{V}^{\dagger}F\tilde{V}\,.
\end{equation}
The transformed Fock matrix $\tilde{F}$ is then diagonalized to obtain a semicanonical representation of the truncated space,
\begin{equation}
    \tilde{F}\tilde{Z}=\tilde{Z}\tilde{\epsilon }\,.
\end{equation}
The eigenvectors $\tilde{Z}$, combined with the retained natural spinors $\tilde{V}$, yield the overall transformation matrix $\tilde{B}$ that relates the canonical and truncated FNS bases,
\begin{equation}
\label{eqn30}
    \tilde{B}=\tilde{V}\tilde{Z}\;.
\end{equation}

Although these natural spinors are appropriate for capturing ground-state correlation and yield reliable ionization energies\cite{chamoli2025reduced,surjuse2022lowcost} and double ionization potentials,\cite{mandal2026thirdorder, mukhopadhyay2026reducedcost} they may not be optimal for describing doubly electron-attached states, as observed in electron-attachment\cite{mukhopadhyay2025reducedcost} and excited-state studies.\cite{manna2025reduced, chakraborty2025low,mukhopadhyay2025reducedcost} This limitation arises because the electron density can change significantly from one excited state to another. A more balanced description can be achieved by improving the density matrix through the inclusion of state-specific information obtained from lower-level methods. This idea forms the basis of the state-specific natural spinor scheme.  The method was first introduced by Helmich and Hättig,\cite{helmich2011locala} and further developed by Kállay and co-workers\cite{mester2017reducedcosta} for excitation energy calculations in the coupled-cluster framework.

Adapting the framework of Head-Gordon and co-workers,\cite{head-gordon1999quasidegenerate} one can construct the DEA-CIS(D$_{\infty}$) Jacobian matrix as
\begin{widetext}
\begin{equation}
\label{eq:cis_d_inf}
A^{\text{DEA-CIS}(D_{\infty})} = \begin{pmatrix}
\bra{\mu_1} [(\hat{H} + [\hat{H}, \hat{T}_2^{(1)}]), \hat{\tau}_{\nu_1}] \ket{HF} & \bra{\mu_1} [\hat{H}, \hat{\tau}_{\nu_2}] \ket{HF} \\
\bra{\mu_2} [\hat{H}, \hat{\tau}_{\nu_1}] \ket{HF} & \delta_{\mu_2 \nu_2} \epsilon_{\mu_2}
\end{pmatrix}\,.
\end{equation}
\end{widetext}
Here, $\hat{H}$ denotes the normal ordered Hamiltonian, $\hat{T}_2^{(1)}$ is the first-order (MP2) double-excitation operator, and $\epsilon_{\mu_2} = \epsilon_a + \epsilon_b + \epsilon_c - \epsilon_i$ is the zeroth-order energy of a $3p1h$ configuration. The excitation-class operators are $\hat{\tau}_{\nu_1} = \hat{a}^\dagger_{a}\hat{a}^\dagger_{b}$ and $\hat{\tau}_{\nu_2} = \hat{a}^\dagger_{a} \hat{a}^\dagger_{b} \hat{a}^\dagger_{c} \hat{a}_{i}$. Symmetrization of Eq.~\eqref{eq:cis_d_inf} yields the DEA-ADC(2) matrix\cite{hattig2000cc2,hattig2005structure,winter2011scaled}
\begin{equation}
A^{\text{DEA-ADC}(2)} = \frac{1}{2} \left( A^{\text{DEA-CIS}(D_{\infty})} + (A^{\text{DEA-CIS}(D_{\infty})})^{\dagger} \right)\,.
\end{equation}
Since the doubles--doubles block $A_{\mu_2\nu_2}$ is diagonal, the doubles component of the excitation vector can be written as
\begin{equation}
\label{eq:doubles_vec}
R_{\mu_2} = - \sum_{\nu_1} \frac{A_{\mu_2 \nu_1} R_{\nu_1}}{\epsilon_{\mu_2} - \bar{\omega}}\,.
\end{equation}
As a result, the ADC(2) eigenvalue problem reduces to an effective nonlinear eigenvalue equation in the single-excitation space, given by\cite{hattig2000cc2,hattig2005structure,winter2011scaled}
\begin{equation}
\label{eq:dea-cis_d_big_eq}
\begin{split}
\sum_{\nu_1} \left[ A_{\mu_1 \nu_1} - \sum_{\gamma_2} \frac{A_{\mu_1 \gamma_2} A_{\gamma_2 \nu_1}}{\epsilon_{\gamma_2} - \bar{\omega}} \right] R_{\nu_1} 
&= \sum_{\nu_1} A_{\mu_1 \nu_1}^{\text{eff}}(\bar{\omega}) R_{\nu_1} \\
&= \bar{\omega} R_{\mu_1},
\end{split}
\end{equation}
where $\bar{\omega}$ is the DEA-ADC(2) energy, and the doubles part of the excitation vector, $R_{\mu_2}$, is recovered from the singles amplitudes $R_{\nu_1}$ through Eq.~\eqref{eq:doubles_vec}. The DEA-CIS(D) method is then obtained from a non-iterative approximation to Eq.~\eqref{eq:dea-cis_d_big_eq},
\begin{equation}
\omega_{\text{CIS}(D)} = \omega_{\text{CIS}} + R_{\text{CIS}}^{\dagger} A_{\text{SD}} (\omega_{\text{CIS}} I - \epsilon_{\text{D}})^{-1} A_{\text{DS}} R_{\text{CIS}}\,,
\end{equation}
where $\epsilon_{\text{D}}$ is the diagonal matrix of $3p1h$ zeroth-order energies $\epsilon_{\mu_2}$, and $A_{\text{SD}}$ ($A_{\text{DS}}$) is the singles--doubles (doubles--singles) block of the Jacobian. In the present work, the SS-FNS basis is constructed using the computationally less demanding DEA-CIS(D) approach instead of DEA-ADC(2). Figure S1 of the Supplementary Material presents a comparison of the DEA-CIS(D) and DEA-ADC(2)\cite{thielen2023development} based SS-FNS-CD-X2CAMF-DEA-EOM-CCSD approaches, showing very similar behavior for the test case considered. Nevertheless, we emphasize that DEA-ADC(2) natural spinors are expected to be more accurate owing to their more complete description of the underlying wavefunction. For the $\mu$th target state, the virtual–virtual block of the state-specific one-particle density matrix takes the form
\begin{equation}
\label{eq:dss}
    D_{ab}^{\text{SS}}(\mu)=D_{ab}^{\text{MP2}}+D_{ab}^{\text{DEA-CIS(D)}}(\mu)\,,
\end{equation}\\
where $D_{ab}^{\text{MP2}}$ and $D_{ab}^{\text{DEA-CIS(D)}}(\mu)$ are the virtual--virtual blocks of the one-particle reduced density matrix from MP2 and DEA-CIS(D), respectively. The SS-FNS spinors are obtained by diagonalizing $D^{\text{SS}}(\mu)$ exactly as the ground-state FNS are obtained from $D^{\text{MP2}}_{ab}$.

For EOM-CCSD calculations, the computational cost of the ground-state CCSD step scales as $\mathcal{O}(N_o^2 N_v^4)$, where $N_o$ and $N_v$ denote the number of occupied and virtual spinors, respectively. In the conventional SS-FNS approach, separate CCSD calculations are required for each target state, which significantly increases the overall computational expense. To address this limitation, we employ a projected formulation of the SS-FNS scheme. In this approach, the CCSD calculation is performed only once in the truncated ground-state FNS basis. The resulting amplitudes are then projected and transformed for each DEA state into the corresponding truncated SS-FNS basis. Let $\tilde{B}_{gr}$ denote the transformation matrix from the canonical basis to the ground-state FNS basis, and $\tilde{B}_{\mu}$ the corresponding transformation matrix to the truncated SS-FNS basis of the $\mu$th DEA state. The CCSD amplitudes are then carried from the ground-state FNS basis to the $\mu$th DEA state's SS-FNS basis by
\begin{equation}
    \tilde{B}_{gr\to\mu} = \tilde{B}_{\mu}^{\dagger}\tilde{B}_{gr}\,.
\end{equation}
The EOM calculation is subsequently performed in the truncated SS-FNS basis of the respective state. A perturbative correction can be added to the DEA energy obtained in the truncated SS-FNS basis to compensate for the truncation,
\begin{equation}
\begin{aligned}
        \omega_{\text{EOM}}^{\text{corrected}}(\mu) &= \omega_{\text{EOM}}^{\text{uncorrected}}(\mu) \\
        & +\omega_{\text{DEA-CIS(D)}}^{\text{canonical}}(\mu)-\omega_{\text{DEA-CIS(D)}}^{\text{SS-FNS}}(\mu)\,.
\end{aligned}
\end{equation}

\subsection{Cholesky Decomposition}
\label{sec:theory_cholesky}
The Cholesky decomposition (CD) technique, first introduced by Beebe and Linderberg,\cite{beebe1977simplifications} provides a computationally attractive means of approximating the positive semi-definite ERI tensor as a product of lower-triangular matrices, commonly referred to as Cholesky vectors (CVs). Within this framework, the ERI tensor in the atomic-orbital (AO) basis is expressed as
\begin{equation}
\label{eqn:Cholesky}
    \left(\mu\nu|\kappa\lambda\right) \approx \sum_{P=1}^{n_{\text{CD}}}{L_{\mu\nu}^{P}L_{\kappa\lambda}^{P}}\,.
\end{equation}
Here, $L^{P}$ are the CVs, $\mu,\nu,\kappa,\lambda$ are AO indices, and $n_{\text{CD}}$ is the number of CVs. In the decomposition procedure, the largest diagonal elements of the ERI matrix are identified iteratively, and the corresponding CVs are generated until the maximum diagonal element becomes smaller than the chosen Cholesky threshold ($\tau$), which determines the overall accuracy.\cite{aquilante2011cholesky,folkestad2019efficient,zhang2021minimal} The CVs obtained in the AO basis are then transformed to the molecular-spinor basis by contraction with the AO--MO coefficient matrix from the SCF calculation,
\begin{equation}
\label{eqn34}
    L_{pq}^{P} = \sum_{\mu\nu}C_{\mu p}^{*} L_{\mu\nu}^{P}C_{\nu q}\,.
\end{equation}
This significantly enhances computational efficiency, since the explicit construction of several demanding integrals, especially those involving three and four virtual indices, is completely circumvented; instead, the corresponding terms are assembled from the CVs on the fly.

\section{\label{sec:compu_details}Computational details}
The canonical, FNS, and SS-FNS variants of the DEA-EOM-CCSD($3p1h$) method have been implemented in our in-house quantum chemistry software \texttt{BAGH}.\cite{dutta2025bagh} The code is primarily written in Python, while the computationally demanding components are handled using C++, Fortran, and Cython. For the evaluation of one- and two-electron integrals, \texttt{BAGH} interfaces with external packages such as \texttt{PySCF},\cite{sun2015libcint,sun2018pyscf,sun2020recent} \texttt{GAMESS-US},\cite{barca2020recent} \texttt{DIRAC},\cite{jensen2022dirac} and \texttt{socutils}.\cite{wang2025socutils} At present, \texttt{BAGH} supports energy and property calculations for a wide range of non-relativistic and relativistic (four- and two-component) methods based on many-body perturbation theory, coupled-cluster, and algebraic diagrammatic construction approaches. All calculations reported in this work were performed using the \texttt{PySCF} interface, where the X2CAMF-HF step was carried out using the \texttt{socutils} package. Cholesky vectors of the LOO, LOV, and LVV types were generated in the canonical basis and used in the canonical DEA-CIS(D) calculations wherever necessary. In this notation, the leading L denotes the CVs index, and O and V denote occupied and virtual spinors, respectively. The LOV and LVV integrals were then transformed into the SS-FNS basis in an on-the-fly manner. This approach not only improves the efficiency of the integral transformation step but also substantially reduces the overall memory footprint. In all calculations, a CD threshold of $10^{-5}$ was adopted.\cite{chamoli2025frozen,chamoli2025reduced} Four-component relativistic calculations were carried out without the CD approximation for the two-electron integrals. The dyall.v$x$z ($x=2,3,4$) hierarchy of basis sets was employed.\cite{dyall1998relativistic,dyall2002relativistic,dyall2006relativistic,dyall2016relativistic} Augmented versions of the dyall.v4z basis set were generated using the \texttt{DIRAC} software package. For GaH and InH, the experimental bond lengths of 1.663 and 1.838 \AA\ were used, respectively.\cite{huber1979constantsa} The experimental bond lengths of 2.166 and 2.557 \AA\ were adopted for Se$_2$ and Te$_2$, respectively.\cite{huber1979constantsa} For Po$_2$, a bond length of 2.795 \AA\ reported in Ref. \onlinecite{rota2011zero} was used. In all correlation calculations reported in this work, the frozen-core approximation was adopted. For the statistical error analysis, MAE and MAD denote the mean absolute error and the maximum absolute deviation, respectively.
\begin{figure*}
    \centering
    \includegraphics[width=\textwidth]{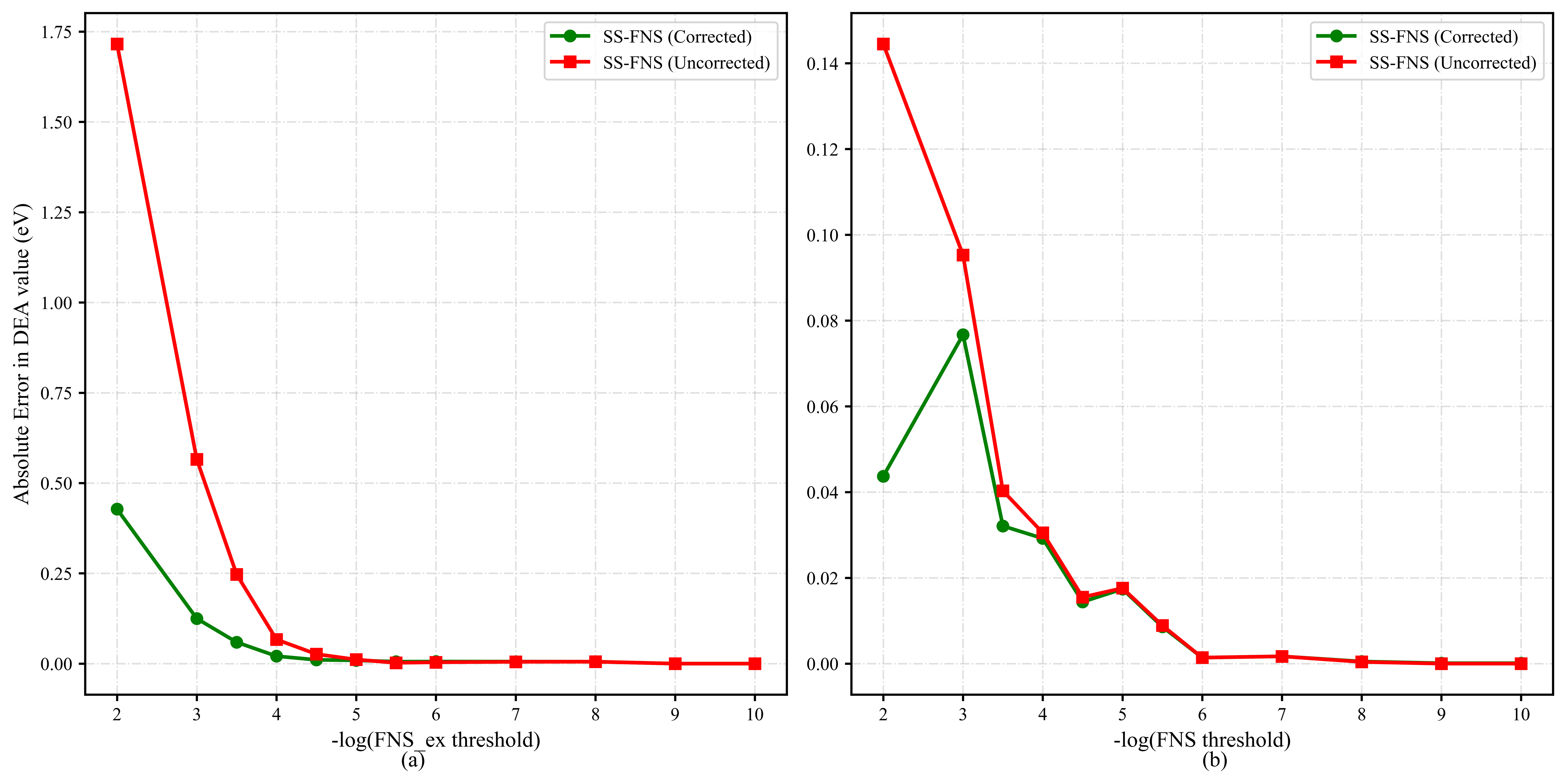}
    \caption{\label{fig:comb_fns} Convergence of the absolute error in the DEA value (in eV) obtained using the SS-FNS-CD-X2CAMF-DEA-EOM-CCSD method for the InH molecule with the s-aug-dyall.v2z basis set.
(a) Variation of the error with respect to the excited-state FNS threshold, where the untruncated canonical result is used as the reference.
(b) Variation with respect to the ground-state FNS threshold, with the excited-state threshold fixed at $10^{-4.5}$; the corresponding value at this threshold is taken as the reference.}
\end{figure*}
To estimate the complete basis set (CBS) limit for the singlet--triplet gap in dihalocarbenes, a two-point extrapolation was performed using the dyall.v3z and dyall.v4z results. The HF energy was extrapolated with the exponential expression\cite{feller1993use}
\begin{equation}
E_{\mathrm{HF}}(X)=E_{\mathrm{HF}}^{\mathrm{CBS}}+Ae^{-\beta X},
\end{equation}
using $\beta=1.62$.~\cite{pansini2016extrapolation} For the correlation energy, the inverse-cubic extrapolation~\cite{martin1996initio}
\begin{equation}
E_{\mathrm{corr}}(X)=E_{\mathrm{corr}}^{\mathrm{CBS}}+\frac{A}{X^3}
\end{equation}
was employed. The corresponding CBS energy was obtained by combining the extrapolated HF and correlation contributions as $E_{\mathrm{CBS}}=E_{\mathrm{HF}}^{\mathrm{CBS}}+E_{\mathrm{corr}}^{\mathrm{CBS}}$.

\section{\label{sec:results}Results and discussion}
\subsection{\label{subsec:choice_of_fns_thresh}Selection of optimal truncation thresholds}
The selection of an appropriate truncation threshold is essential
\begin{table}
\caption{\label{tab:zn-gah-compare}
Comparison of DIP and EE values (in eV) for Zn and GaH, calculated using the four-component (4c) and CD-X2CAMF implementations of the SS-FNS-DEA-EOM-CCSD method with the dyall.v4z basis set.}
\begin{ruledtabular}
\begin{tabular}{ccccc}
Atom/Molecule & State & 4c & CD-X2CAMF & Expt.\cite{kramida2024nist,huber1979constantsa} \\
\hline
\multirow{5}{*}{Zn}
& $^1S_0$ & 27.114 & 27.107 & 27.359 \\
& $^3P_0$ & 3.963  & 3.961  & 4.006  \\
& $^3P_1$ & 3.986  & 3.984  & 4.030  \\
& $^3P_2$ & 4.034  & 4.031  & 4.078  \\
& $^1P_1$ & 5.794  & 5.791  & 5.796  \\
\hline
\multirow{5}{*}{GaH}
& $\Sigma_0^+$ & 25.814 & 25.808 & --    \\
& $\Pi_0^-$    & 2.080  & 2.075  & 2.149 \\
& $\Pi_0^+$    & 2.081  & 2.077  & 2.150 \\
& $\Pi_1$      & 2.114  & 2.107  & 2.185 \\
& $\Pi_2$      & 2.165  & 2.159  & --    \\
\end{tabular}
\end{ruledtabular}
\end{table}
for balancing computational efficiency and accuracy in FNS-based approaches. In our previous studies,\cite{manna2025reduced,mandal2026thirdorder,mukhopadhyay2026reducedcosta} we have shown that truncation applied directly in the canonical virtual space leads to slow convergence toward the untruncated canonical results. In contrast, truncation in the FNS virtual space substantially accelerates convergence. In this section, we examine the convergence of the FNS and SS-FNS based DEA-EOM-CCSD methods for different occupation number based truncation thresholds. For this analysis, we considered InH in its 2+ cationic reference state and computed the ground-state energy of neutral InH using the s-aug-dyall.v2z basis set with canonical, FNS-, and SS-FNS-based DEA-EOM-CCSD methods. Figure S2 of the Supplementary Material shows the absolute error of the FNS-DEA-EOM-CCSD method relative to the untruncated canonical value as a function of the truncation 
\begin{table}
\caption{\label{tab:basis-conv}
Basis set convergence of SS-FNS-CD-X2CAMF-DEA-EOM-CCSD values (in eV) for GaH obtained with different dyall basis sets. The tabulated values are DEA energies relative to the GaH$^{2+}$ reference; the $\Sigma_0^+$ entry is the negative of the DIP reported in Table~\ref{tab:zn-gah-compare}.}
\begin{ruledtabular}
\begin{tabular}{ccccccc}
State & \makecell{dyall.\\v2z} &
       \makecell{dyall.\\v3z} &
       \makecell{dyall.\\v4z} &
       \makecell{s-aug-\\dyall.\\v4z} &
       \makecell{d-aug-\\dyall.\\v4z} &
       \makecell{t-aug-\\dyall.\\v4z} \\
\hline
$\Sigma_0^+$ & -25.43 & -25.68 & -25.81 & -25.81 & -25.81 & -25.81 \\
$\Pi_0^-$    & -23.40 & -23.63 & -23.73 & -23.74 & -23.74 & -23.74 \\
$\Pi_0^+$    & -23.39 & -23.63 & -23.73 & -23.74 & -23.74 & -23.74 \\
$\Pi_1$      & -23.36 & -23.59 & -23.70 & -23.70 & -23.70 & -23.71 \\
$\Pi_2$      & -23.32 & -23.56 & -23.65 & -23.65 & -23.66 & -23.66 \\
\end{tabular}
\end{ruledtabular}
\end{table}
threshold. Even at a threshold of 10$^{-8}$, which truncates the virtual space by 45\%, the absolute error remains large, at approximately 0.1 eV. In Figure \ref{fig:comb_fns}(a), the same error is shown for the SS-FNS-DEA-EOM-CCSD method as a function of the excited-state FNS threshold, while keeping the ground-state FNS threshold fixed at zero. In contrast to the FNS-based approach, the SS-FNS method (with perturbative correction added) yields an absolute error of only 0.01 eV at a threshold of 10$^{-4.5}$, despite truncating the virtual space by as much as 73\%. As shown in Figure \ref{fig:comb_fns}(a), the inclusion of the perturbative correction leads to faster convergence. For example, at an excited-state FNS threshold of 10$^{-4.5}$, the absolute error decreases from 0.03 eV in the uncorrected calculation to 0.01 eV when the perturbative correction is included. Accordingly, only the corrected DEA values are reported hereafter. Finally, with the excited-state FNS threshold fixed at 10$^{-4.5}$, we varied the ground-state FNS threshold and plotted the resulting absolute errors in Figure \ref{fig:comb_fns}(b). Considering the associated computational cost, we selected 10$^{-4.5}$ as the optimal ground state FNS threshold as well, for which the corrected absolute error remains small at only 0.014 eV. Therefore, all subsequent calculations in this work employ a value of 10$^{-4.5}$ for both the ground and excited state FNS thresholds, unless stated otherwise.

\begin{table*}
\caption{\label{tab:zn-cd-hg}
Comparison of DIP and EEs (in eV) of Zn, Cd, and Hg atoms computed using the SS-FNS-CD-X2CAMF-DEA-EOM-CCSD method with the dyall.v4z basis set, alongside available theoretical and experimental data. The values in parentheses correspond to energy splittings measured relative to the $J$-averaged triplet energy.}
\begin{ruledtabular}
\begin{tabular}{ccccc}
Atom & State & Previous calculations\footnotemark[1] & This work & Experiment\cite{kramida2024nist} \\
\hline
Zn & $^1S_0$ & 27.102, 27.026 & 27.107 & 27.359 \\
   & $^3P_0$ &  3.944, 3.896, 3.944 &  3.961 (-0.047) &  4.006 (-0.048) \\
   & $^3P_1$ &  3.967, 3.918, 3.969 &  3.984 (-0.024) &  4.030 (-0.024)\\
   & $^3P_2$ &  4.014, 3.963, 4.020 &  4.031 (0.023)  &  4.078 (0.024)\\
   & $^1P_1$ &  5.755, 5.792, 5.786 &  5.791 (1.783)  &  5.796 (1.742)\\
\hline
Cd & $^1S_0$ & 25.762, 25.726 & 25.765 & 25.902 \\
   & $^3P_0$ &  3.699, 3.660, 3.624 &  3.692 (-0.141) &  3.734 (-0.140)\\
   & $^3P_1$ &  3.765, 3.726, 3.688 &  3.759 (-0.074) &  3.801 (-0.073)\\
   & $^3P_2$ &  3.906, 3.868, 3.825 &  3.905 (0.072) &  3.946 (0.072)\\
   & $^1P_1$ &  5.384, 5.474, 5.286 &  5.403 (1.570) &  5.417 (1.543)\\
\hline
Hg & $^1S_0$ & 28.961, 29.020 & 29.004 & 29.194 \\
   & $^3P_0$ &  4.615, 4.499, 4.791 &  4.605 (-0.522) &  4.667 (-0.514)\\
   & $^3P_1$ &  4.832, 4.732, 4.971 &  4.828 (-0.299) &  4.886 (-0.295)\\
   & $^3P_2$ &  5.398, 5.351, 5.407 &  5.410 (0.283)  &  5.461 (0.280)\\
   & $^1P_1$ &  6.629, 6.721, 6.720 &  6.669 (1.542) &  6.704 (1.523)\\
\end{tabular}
\end{ruledtabular}
\footnotetext[1]{For the DIP ($^1S_0$ state), the first and second entries correspond to DEA-EOM-SOC\cite{guo2020equation} and DIP-EOM-SOC\cite{wang2015equationofmotion} results, respectively. For EEs, the first, second, and third entries are taken from DEA-EOM-SOC,\cite{guo2020equation} EOM-SOC-CCSD,\cite{wang2014equation} and EOM-SOC-CCSDT calculations,\cite{akinaga2017two} respectively.}
\end{table*}

\subsection{\label{subsec:cd_vs_4c} Validation of CD-X2CAMF results against 4c SS-FNS-DEA-EOM-CCSD}
As a benchmark of the CD-X2CAMF-based SS-FNS-DEA-EOM-CCSD approach, we evaluated the lowest DIP and the excitation energies of the four lowest excited states of Zn and GaH and compared them with the corresponding four-component results. All calculations were performed using the dyall.v4z basis set, and the results are summarized in Table \ref{tab:zn-gah-compare}. It is evident that, for all states considered, the deviation between the four-component and CD-X2CAMF-based results does not exceed 0.007 eV. This demonstrates that the CD-X2CAMF-based SS-FNS-DEA-EOM-CCSD method can achieve near four-component accuracy, while providing substantial reductions in both computational cost and memory requirements. Consequently, all subsequent calculations in this work are carried out using the SS-FNS-CD-X2CAMF-DEA-EOM-CCSD method.

\subsection{\label{subsec:basis_set_conv} Basis set convergence}
The choice of basis set plays a crucial role in the calculation of DEA energies. To assess the convergence of the DEA values with respect to basis set size, we employed the widely used dyall basis set hierarchy, dyall.v$x$z ($x$ = 2, 3, 4).\cite{dyall1998relativistic,dyall2002relativistic,dyall2006relativistic,dyall2016relativistic} In addition, to investigate the effect of diffuse functions, we studied single-, double-, and triple-augmented variants of the dyall.v4z basis and analyzed their impact on the computed DEA values. These calculations were carried out for the five lowest double-electron-attachment energies with respect to the GaH$^{2+}$ reference state. The corresponding results are presented in Table~\ref{tab:basis-conv}. An average lowering of approximately -0.24 eV in the DEA values is observed across all five states when the basis set is increased from dyall.v2z to dyall.v3z. This average lowering decreases to -0.11 eV upon further expansion from dyall.v3z to dyall.v4z. Convergence beyond this level cannot be assessed due to the absence of the dyall.v5z basis set. Nevertheless, single, double, and triple augmentation of the dyall.v4z basis leads to only marginal variations, with DEA values changing by at most -0.01 eV or remaining effectively unchanged. Consequently, dyall.v4z was selected as the working basis set for all subsequent calculations.

\subsection{\label{subsec:atoms} Excitation energies of atoms}
In this section, we benchmark the performance of our method by computing the lowest DIP and EEs for several low-lying excited states of selected heavy atoms. As a first test set, we consider the group 12 elements Zn, Cd, and Hg. Table \ref{tab:zn-cd-hg} reports the corrected SS-FNS-CD-X2CAMF-DEA-EOM-CCSD results, along with comparisons to available theoretical and experimental data. The $^1S_0$ state corresponds to the DIP of the atom, while the remaining states originate from the $ns^1np^1$ excited state electronic configurations. Given the large DIP range of 25--30 eV, our calculated values show reliable agreement with experiment, the largest deviation being $-0.25$ eV for Zn. Moreover, the calculated DIPs are systematically underestimated, consistent with the trend observed in previous theoretical studies.\cite{guo2020equation,wang2015equationofmotion} With respect to experimental data, the EEs of Zn, Cd, and Hg obtained from previously reported theoretical approaches exhibit MAE/MAD values of 0.052/0.075 eV for DEA-EOM-SOC,\cite{guo2020equation} 0.090/0.168 eV for EOM-SOC-CCSD,\cite{wang2014equation} and 0.079/0.131 eV for EOM-SOC-CCSDT.\cite{akinaga2017two} In comparison, the present method yields smaller errors, with an MAE of 0.041 eV and a MAD of 0.062 eV; the largest deviation occurs for the $^3P_0$ state of Hg. In Table \ref{tab:zn-cd-hg}, we also report the energy splittings of the excited states relative to the $J$-averaged triplet reference. The splittings obtained for the $^3P_0$, $^3P_1$, and $^3P_2$ states are in excellent agreement with experimental findings. In contrast, the splitting associated with the $^1P_1$ state shows a somewhat larger discrepancy, consistent with earlier theoretical reports.\cite{guo2020equation,wang2014equation,akinaga2017two}

In the second test set, we examined the group 14 elements Ge, Sn, and Pb, whose electronic excitations have been comparatively less explored in the literature. Using the $(n-1)d^{10}ns^2$ closed-shell configuration as the reference, we evaluated the lowest DIP ($^3P_0$ state) and the low-lying excited-state energies of the $np^2$ manifold. The computed values are listed in Table \ref{tab:ge-sn-pb} together with available DEA-EOM-SOC results and experimental measurements for comparison. For the dicationic reference states in the SS-FNS-DEA-EOM-CCSD calculations, two frozen-core schemes were considered: (a) 20 correlated electrons for each atom; and (b) 28, 30, and 44 correlated electrons for Ge, Sn, and Pb, respectively, with orbitals up to the [He], [Ar], and [Kr] closed-shell configurations frozen, respectively. The extended correlation space systematically improves the DIP values, reducing the errors from -0.145 to -0.108 eV for Ge, from -0.171 to -0.157 eV for Sn, and from -0.185 to -0.158 eV for Pb. The excitation energies also show an overall improvement upon increasing the correlation space. With the first frozen-core scheme, in which 20 occupied spinors are correlated for each atom, the MAE and MAD with respect to experiment are 0.037 and 0.129 eV, respectively, with the largest deviation occurring for the $^1S_0$ state of Pb. Upon inclusion of the extended correlation space, these values decrease to 0.028 and 0.085 eV, respectively. These results demonstrate that correlating a larger number of electrons improves the agreement with experiment for Ge, Sn, and Pb. In comparison, DEA-EOM-SOC\cite{guo2020equation} gives MAE and MAD values of 0.028~eV and 0.094~eV, respectively, for the excitation energies, with 20 electrons correlated in each atom.
\begin{table}
\caption{\label{tab:ge-sn-pb}
Comparison of DIP and EEs (in eV) of Ge, Sn, and Pb atoms computed using the SS-FNS-CD-X2CAMF-DEA-EOM-CCSD method with the dyall.v4z basis set, using the corresponding dicationic reference states, alongside available theoretical and experimental data. In scheme (a), 20 electrons were correlated for each atom. In scheme (b), 28, 30, and 44 electrons were correlated for Ge, Sn, and Pb, respectively.}
\begin{ruledtabular}
\begin{tabular}{cccccc}
Atom & State & DEA-EOM-SOC\cite{guo2020equation} & \multicolumn{2}{c}{This work} & Experiment\cite{kramida2024nist} \\
     &        &             & (a)  & (b) & \\
\hline
Ge & $^3P_0$ & 23.709 & 23.689 & 23.726 & 23.834 \\
   & $^3P_1$ &  0.067 &  0.074 & 0.073 &  0.069 \\
   & $^3P_2$ &  0.172 &  0.176 & 0.178 &  0.175 \\
   & $^1D_2$ &  0.896 &  0.912 & 0.909 &  0.883 \\
   & $^1S_0$ &  2.038 &  2.065 & 2.056 & 2.029 \\
\hline
Sn & $^3P_0$ & 21.907 & 21.806 & 21.820 & 21.977 \\
   & $^3P_1$ &  0.201 &  0.209 & 0.209 & 0.210 \\
   & $^3P_2$ &  0.413 &  0.416 & 0.419 & 0.425 \\
   & $^1D_2$ &  1.061 &  1.080 & 1.076 & 1.068 \\
   & $^1S_0$ &  2.119 &  2.148 & 2.132 & 2.128 \\
\hline
Pb & $^3P_0$ & 22.292 & 22.264 & 22.291 & 22.449 \\
   & $^3P_1$ &  0.922 &  0.910 & 0.912 & 0.969 \\
   & $^3P_2$ &  1.274 &  1.268 & 1.268 & 1.320 \\
   & $^1D_2$ &  2.566 &  2.565 & 2.575 & 2.660 \\
   & $^1S_0$ &  3.568 &  3.524 & 3.588 & 3.653 \\
\end{tabular}
\end{ruledtabular}
\end{table}

\subsection{\label{subsec:molecules} Excitation energies of diatomic molecules}
The chalcogen homonuclear dimers are open-shell systems characterized by two electrons occupying the $\pi^*_g$ molecular orbital. Starting from a dicationic closed-shell reference state, the DEA-EOM-CCSD method provides a suitable framework for describing both the ground and excited states of these molecules. In this work, we have investigated the zero-field splitting (ZFS) between the $X_2\,1$ and $X_1\,0^+$ states, which originate from the spin-orbit free $X\,^3\Sigma^-$ state. In addition, we have computed the vertical excitation energies of the $a\,2$ and $b\,0^+$ states relative to the ground $X_1\,0^+$ state. Bolvin and co-workers have performed an extensive study\cite{rota2011zero} of these systems using high-level two- and four-component relativistic methods, including intermediate Hamiltonian Fock–space coupled-cluster (IHFSCC) and general active space configuration interaction (GASCI), along with two-step approaches such as spin–orbit complete active space second-order perturbation theory (SO-CASPT2). The results obtained in the present work are compared in Table \ref{tab:group16-dimers} with these earlier theoretical studies, along with available experimental and DEA-EOM-SOC data.\cite{guo2020equation} In all cases, 30 electrons were included in the correlation treatment. For Po$_2$, at a CD threshold of $10^{-5}$, a total of 3248 Cholesky vectors were generated. The canonical virtual space consisted of 1190 spinors. In the excited-state calculations, the SS-FNS basis led to a substantial reduction of the virtual space, with approximately 80\% of the spinors being truncated on average. It can be seen from Table~\ref{tab:group16-dimers} that, for the selenium and tellurium dimers, the ZFS of the $X\,^3\Sigma^-$ state is slightly underestimated, whereas the excitation energy of the $b\,0^+$ state is overestimated relative to the experimental values. Unfortunately, no experimental data are currently available for Po$_2$. Nevertheless, our results are in reasonable agreement with previous theoretical studies,\cite{rota2011zero} the largest discrepancy being for the ZFS of Po$_2$. It is also evident that the ZFS increases along the series, which can be attributed to the progressively stronger spin-orbit coupling, as expected.
\begin{table}
\caption{\label{tab:group16-dimers}
Comparison of calculated zero-field splittings ($X_2\,1$) and vertical excitation energies ($a\,2$ and $b\,0^+$) (in cm$^{-1}$) for chalcogen homonuclear diatomic molecules with available theoretical and experimental data.}
\begin{ruledtabular}
\begin{tabular}{ccccc}
Molecule & State & Previous calculations\footnotemark[1] & This work & Experiment \\
\hline
Se$_2$ & $X_2\,1$ & 574, 446 & 499 & 510\cite{huber1979constantsa} \\
      & $a\,2$    & 4242, 3470 & 4182 & -- \\
      & $b\,0^+$   & 7939, 7463 & 8125 & 
\makecell[c]{7417\cite{winter1980b1s}, 7936\cite{winter1982b1s}\\
, 7958\cite{prosser1980nature}} \\
\hline
Te$_2$ & $X_2\,1$ &  2178,  1793,  1724 & 1811 & 1975\cite{huber1979constantsa} \\
      & $a\,2$    &  5029,  4464,  4875 & 4793 & -- \\
      & $b\,0^+$  &  9724,  8756,  9767 & 9667 & 9600\cite{effantin1980new} \\
\hline
Po$_2$ & $X_2\,1$ & 7470, 6903 & 5851 & -- \\
      & $a\,2$    & 9665, 7619 & 7855 & -- \\
      & $b\,0^+$  & 16864, 16271 & 16096 & -- \\
\end{tabular}
\end{ruledtabular}
\footnotetext[1]{The first and second entries in this column report IHFSCC and SO-CASPT2 results,\cite{rota2011zero} respectively. For Te$_2$, a third entry provides DEA-EOM-SOC\cite{guo2020equation} data.}
\end{table}

To further assess the applicability of the present method, we consider the group-13 hydrides GaH, InH, and TlH. Starting from the $(1\sigma)^2(2\sigma)^0(1\pi)^0$ reference state, we have studied the ground state $(1\sigma)^2(2\sigma)^2(1\pi)^0$ as well as the excited states arising from the $(1\sigma)^2(2\sigma)^1(1\pi)^1$ electronic configuration. For these states, we have computed key spectroscopic constants, including equilibrium bond lengths, harmonic vibrational frequencies, and adiabatic excitation energies. In all calculations, a total of 20 electrons were correlated in each of these molecules. The computed results are summarized in Table~\ref{tab:dea-diatomics-updated}, along with DEA-EOM-SOC\cite{guo2020equation} and experimental data wherever available. We observe that the present method predicts shorter bond lengths than experiment, with a corresponding overestimation of the harmonic frequencies. This behavior is consistent with earlier DEA-EOM-SOC calculations. The adiabatic excitation energies are also in reasonable agreement with the previous theoretical results and, where available, with experiment.
\begin{table*}
\caption{\label{tab:dea-diatomics-updated}
Comparison of diatomic spectroscopic constants calculated using the SS-FNS-CD-X2CAMF-DEA-EOM-CCSD method with the dyall.v4z basis set, together with available theoretical and experimental data for GaH, InH, and TlH. The reported values are bond lengths $R_e$ (\AA), harmonic frequencies $\omega_e$ (cm$^{-1}$), and adiabatic excitation energies $T_e$ (eV).}
\begin{ruledtabular}
\begin{tabular}{llcccccccccccccc}
Molecule & Remarks\footnotemark[1]
& \multicolumn{2}{c}{$\Sigma_0^+$}
& \multicolumn{3}{c}{$\Pi_0^-$}
& \multicolumn{3}{c}{$\Pi_0^+$}
& \multicolumn{3}{c}{$\Pi_1$}
& \multicolumn{3}{c}{$\Pi_2$} \\
\cline{3-4} \cline{5-7} \cline{8-10} \cline{11-13} \cline{14-16}
& 
& $R_e$ & $\omega_e$
& $R_e$ & $\omega_e$ & $T_e$
& $R_e$ & $\omega_e$ & $T_e$
& $R_e$ & $\omega_e$ & $T_e$
& $R_e$ & $\omega_e$ & $T_e$ \\
\hline
\multirow{3}{*}{GaH}
& This work
& 1.647 & 1660
& 1.577 & 1725 & 2.057
& 1.580 & 1752 & 2.058
& 1.580 & 1776 & 2.089
& 1.578 & 1707 & 2.136 \\
& DEA-EOM-SOC
& 1.642 & 1654
& 1.569 & 1775 & 2.058
& 1.569 & 1776 & 2.058
& 1.570 & 1775 & 2.094
& 1.570 & 1773 & 2.131 \\
& Experiment
& 1.663 & 1605
& 1.634 & 1493 & 2.149
& 1.629 & 1641 & 2.150
& 1.593 & 1631 & 2.185
& 1.578 & -- & -- \\
\hline
\multirow{3}{*}{InH}
& This work
& 1.805 & 1581
& 1.733 & 1698 & 1.893
& 1.732 & 1728 & 1.899
& 1.734 & 1658 & 1.983
& 1.732 & 1525 & 2.092 \\
& DEA-EOM-SOC
& 1.810 & 1544
& 1.731 & 1634 & 1.961
& 1.731 & 1639 & 1.966
& 1.732 & 1629 & 2.048
& 1.731 & 1636 & 2.154 \\
& Experiment
& 1.838 & 1476
& 1.776 & 1303 & 2.012
& 1.779 & 1459 & 2.018
& 1.768 & 1415 & 2.100
& 1.753 & 1301 & 2.207 \\
\hline
\multirow{3}{*}{TlH}
& This work
& 1.857 & 1454
& 1.790 & 1248 & 2.102
& 1.777 & 1298 & 2.190
& 1.806 & 1107 & 2.324
& 1.779 & 1438 & 2.773 \\
& DEA-EOM-SOC
& 1.853 & 1413
& 1.787 & 1265 & 2.127
& 1.774 & 1363 & 2.209
& 1.802 & 1155 & 2.308
& 1.771 & 1346 & 2.783 \\
& Experiment
& 1.870 & 1391
& -- & -- & --
& 1.908 & 759 & 2.197
& -- & -- & --
& 3.310 & -- & 3.018 \\
\end{tabular}
\end{ruledtabular}
\footnotetext[1]{DEA-EOM-SOC values are taken from Ref.~\onlinecite{guo2020equation}; experimental values are taken from Ref.~\onlinecite{huber1979constantsa}.}
\end{table*}

\subsection{\label{subsec:st-gap} Singlet-triplet gap in dihalocarbenes}
Carbenes are important reactive intermediates in organic chemistry and have therefore attracted considerable attention from both theoretical and experimental researchers. Methylene, the simplest carbene, is well established to have a triplet ground state, with singlet-triplet (S-T) gap, $\Delta E_{\text{S-T}}\approx-9$~kcal/mol. However, halogen-substituted carbenes generally display positive S-T gaps, indicating singlet ground states. The S-T separations of CCl$_2$, CBr$_2$, and CI$_2$ obtained from several high-level \textit{ab initio} studies are substantially larger than the experimental estimates derived from negative-ion photoelectron spectroscopy, and satisfactory agreement between theory and experiment has been demonstrated only recently, and only for CCl$_2$.\cite{schwartz1999singlettriplet,wren2009photoelectron} Previous studies have employed DEA-EOM-CC-based methods to investigate the S-T gaps of a variety of carbene species, including methylene, trimethylenemethane, and the benzyne isomers.\cite{shen2013doubly, shen2014doubly,shen2021double} Here, we evaluate the applicability of this methodology to heavier halogen-substituted carbenes by computing the S-T gaps of CCl$_2$, CBr$_2$, and CI$_2$. The molecular geometries were taken from Ref.~\onlinecite{hajgato2000triplet}, and the corresponding Cartesian coordinates are provided in the Supplementary Material. For the correlated calculations, the 1s electrons of carbon were frozen, while for the halogen atoms the cores up to [Ne], [Ar], and [Kr] were frozen for CCl$_2$, CBr$_2$, and CI$_2$, respectively. The S-T gaps obtained in the present work are summarized in Table~\ref{tab:st_gap} along with earlier theoretical and experimental data. Since spin--orbit coupling removes the degeneracy of the triplet state, three S-T gap values are reported for each molecule. The reported values were obtained by extrapolation to the complete-basis-set (CBS) limit using energies calculated with the dyall.v3z and dyall.v4z basis sets. No zero-point energy (ZPE) corrections were included. It can be seen from Table~\ref{tab:st_gap} that our calculated S-T gaps, consistent with previous theoretical studies, are substantially larger than the corresponding experimental values. The S-T gap decreases from CCl$_2$ to CI$_2$, suggesting an increase in the diradical character across the series. The calculated S-T gaps are larger than those obtained in earlier \textit{ab initio} studies. In a recent study, Piecuch and co-workers\cite{shen2026double} showed that the S-T gap of trimethylenemethane decreases from 23.92~kcal/mol to approximately 19~kcal/mol when $4p2h$ correlations are included through an active-space treatment (see Table 3 of their work). This suggests that the inclusion of $4p2h$ correlation effects may be essential for achieving quantitative accuracy in S-T gap calculations.
\begin{table}
\caption{\label{tab:st_gap}
Comparison of singlet-triplet (S-T) gaps (kcal mol$^{-1}$) calculated using the SS-FNS-CD-X2CAMF-DEA-EOM-CCSD method in the CBS limit, together with available theoretical and experimental data for selected dihalocarbenes.}
\begin{ruledtabular}
\begin{tabular}{ccc}
System & S-T gap & Remarks \\
\hline
CCl$_2$  & 25.77, 25.78, 25.79 & This work \\
         & 20.10              & CCSD(T)/6-311++G(3df,2p)$^{a}$ \\
         & 17.57              & CASPT2/AE/6-311++G(3df,3pd)$^{a}$ \\
         & 20.3               & UCCSD(T)/cc-pVQZ$^{b}$ \\
         & 23.8               & HFS/NLC$^{c}$ \\
         & $3 \pm 3$, $20.75 \pm 4.6$           & Experiment$^{d,e}$ \\
\hline
CBr$_2$  & 23.47, 23.48, 23.78 & This work \\
         & 17.02              & CCSD(T)/6-311++G(3df,2p)$^{a}$ \\
         & 14.82             & CASPT2/AE/6-311++G(3df,3pd)$^{a}$ \\
         & 17.2              & RCCSD(T)/ECP28MWB$^{b}$ \\
         & 22.4              & HFS/NLC$^{c}$ \\
         & $2 \pm 3$          & Experiment$^{d}$ \\
\hline
CI$_2$   & 18.04, 18.09, 19.30 & This work \\
         & 13.77              & CCSD(T)/6-311++G(3df,2p)$^{a}$ \\
         & 9.42             & CASPT2/AE/6-311++G(3df,3pd)$^{a}$ \\
         & 9.3              & RCCSD(T)/ECP46MDF$^{b}$ \\
         & 16.5              & HFS/NLC$^{c}$ \\
         & $-1 \pm 3$          & Experiment$^{d}$ \\
\end{tabular}
\end{ruledtabular}
\begin{flushleft}
$^{a}$ Ref.~\onlinecite{hajgato2000triplet}; relativistic correction included.\\
$^{b}$ Ref.~\onlinecite{lee2000lowest}; $^{c}$ Ref.~\onlinecite{russo1992geometries}.\\
$^{d}$ Ref.~\onlinecite{schwartz1999singlettriplet}; $^{e}$ Ref.~\onlinecite{wren2009photoelectron}.
\end{flushleft}
\end{table}

\section{\label{sec:conclusion}Conclusion}
We have presented a computationally efficient two-component relativistic implementation of the DEA-EOM-CCSD($3p1h$) method. We have introduced a new SS-FNS scheme based on the DEA-CIS(D) density. By employing two independently controllable truncation thresholds for the ground- and excited-state frozen natural spinor spaces, we have shown that near-canonical accuracy can be recovered using only a reduced virtual space. This leads to a substantial reduction in the computational cost, especially for DEA-EOM-CCSD, where the doubles excitations are of $3p1h$ type. In addition, the combination of the X2CAMF Hamiltonian with Cholesky decomposition offers significant improvements in both computational cost and memory usage, while maintaining accuracy close to that of full four-component relativistic calculations. Reliable DIPs and EEs, including spin-orbit splittings, are obtained with the present method for the group-12 and group-14 heavy atoms. Vertical excitation energies were also reported for the homonuclear chalcogen dimers, where our results are generally in good agreement with previously reported theoretical values. For the group-13 hydrides, namely GaH, InH, and TlH, relatively larger deviations from the experimental data are observed for the equilibrium bond lengths and harmonic frequencies. However, the calculated adiabatic excitation energies are described reasonably well. The singlet-triplet gaps of the dihalocarbenes were calculated to the complete basis set limit and are found to be larger than those reported in previous \textit{ab initio} studies. Further improvements may be achieved by incorporating $4p2h$ excitations. However, the associated $\mathcal{O}(N^8)$ scaling makes this computationally demanding. A possible strategy to overcome this limitation is the use of an active-space treatment of the $4p2h$ block.\cite{shen2013doubly,shen2026double} We plan to explore this approach in future work.

\section*{Acknowledgements}
The authors gratefully acknowledge financial support from IIT Bombay, including the IIT Bombay Seed Grant (Project No. R.D./0517-IRCCSH0-040), CRG (Project No. CRG/2022/005672), and MATRICS (Project No. MTR/2021/000420) projects of DST-SERB; CSIR-India (Project No. 01(3035)/21/EMR-II); DST-Inspire Faculty Fellowship (Project No. DST/INSPIRE/04/2017/001730); and ISRO (Project No. RD/0122-ISROC00-004). The authors also acknowledge the IIT Bombay supercomputing facility and C-DAC resources (Param Smriti, Param Bramha, and Param Rudra) for providing computational time. S.M. gratefully acknowledges support from the Prime Minister’s Research Fellowship (PMRF).


\renewcommand{\refname}{References}
\nocite{apsrev41Control}
\bibliographystyle{aipnum4-1}
\bibliography{reference-dea-eom, reference_zoetero_DIP}



\end{document}